\documentclass{article}
\usepackage{amsmath,graphicx,mlspconf}
\usepackage{xcolor}
\usepackage{amsmath,graphicx,hyperref}

\usepackage{balance}
\usepackage{booktabs}
\usepackage{adjustbox}
\usepackage{array}
\usepackage{bm}
\usepackage{multirow}
\usepackage{algorithmic}
\usepackage{algorithm}
\usepackage{cite}
\usepackage{subcaption}
\usepackage{amssymb,amsfonts}
\usepackage{url,times,booktabs, tabularx}
\usepackage[activate]{microtype}
\usepackage{comment}

\newcommand{\sstitle}[1]{\smallskip\noindent\textbf{#1.\/}}

\toappear{2026 IEEE International Workshop on Machine Learning for Signal Processing, Sep.\ 28-- Oct.\ 1, 2026, Atlanta, USA}

\title{Retain-Free Machine Unlearning for Speech Emotion Recognition}
\name{Zhao Ren$^{1}$, Rathi Adarshi Rammohan$^{1}$, Kevin Scheck$^{1}$, Thanh Tam Nguyen$^{2}$, Sheng Li$^{3,4}$, Tanja Schultz$^{1}$ \thanks{This study is funded by the Deutsche Forschungsgemeinschaft (DFG, German Research Foundation) through the project ``Silent Paralinguistics" (514018165).}}
\address{$^{1}$Cognitive Systems Lab, University of Bremen, Germany\\
$^{2}$Griffith University, Australia \\
$^{3}$Institute of Science Tokyo, Japan \\
$^{4}$Kyoto University, Japan
}

\begin{document}

\maketitle

\begin{abstract}
Speech Emotion Recognition (SER) infers a speaker's emotional state from speech and is increasingly deployed in human-computer interaction, education, and healthcare. Because speech also carries sensitive personal information, speakers may ask that some of their recordings be deleted, which requires removing the influence of those samples from an already trained SER model. Most machine unlearning methods can meet this request only with access to the remaining training data alongside the samples to be forgotten; this is impractical when the remaining data cannot be redistributed or has itself been deleted, and it adds storage and computation as the data grows. To this end, we propose a retain-free unlearning method that updates a pre-trained SER model using only the forget set. Our key idea is to synthesise adversarial samples from the forget set as a surrogate for the unavailable remaining data, and to constrain each parameter update by its estimated importance so that forgetting does not erase general knowledge. The experiments over several emotional-speech corpora and self-supervised backbones show that our method drives forget-set performance down to near chance while retaining much of the model's utility on the remaining and unseen-speaker data, narrowing the gap to methods that rely on the remaining set.
\end{abstract}

\newcommand{\cem}[1]{\textcolor{blue}{cem: #1}}
\newcommand{\eg}{e.g.,\ }
\newcommand{\ie}{i.e.,\ }

\section{Introduction}
\label{sec:intro}
Understanding emotion from speech is central to natural human--machine interaction~\cite{dutta2025llm}. Speech Emotion Recognition (SER) uses machine learning to infer a speaker's emotional state from speech signals~\cite{wani2021comprehensive}. Self-supervised speech models such as Wav2Vec~\cite{baevski2020wav2vec} and HuBERT~\cite{hsu2021hubert} reach strong SER accuracy after fine-tuning, and are increasingly deployed on speech streams from online platforms, mobile devices, and wearables.
This deployment, however, turns stored and reused speech into a privacy liability~\cite{nguyen2022survey}. A single recording can encode sensitive attributes that identify the speaker and expose emotion or mental-health status~\cite{feng2021privacy, ren2019multi}, and such leakage can be exploited: leaked demographic cues (\eg gender) can degrade downstream tasks such as depression detection~\cite{alsenani2024assessing}. A speaker may therefore ask that specific recordings be deleted. Deleting them from storage alone is not enough, however, because an SER model trained on those samples can still retain, and potentially reveal, what it learnt from them. Removing that residual influence from an already deployed model is the goal of machine unlearning for SER.


Instance-level unlearning removes the influence of specific training samples from a pre-trained model, and most methods assume access to the full training set~\cite{bourtoule2021machine}. In practice they need both the data to be forgotten (the \emph{forget set}) and the remaining training data (the \emph{remain set}): the remain set is what keeps accuracy on retained and unseen data from collapsing once the forgetting objective begins to erase the model. This reliance is exactly what breaks in the privacy setting above, since the remain set may be unavailable when users prohibit its redistribution or when it has itself been deleted, and revisiting it is costly in storage and computation as streaming speech accumulates. Forgetting with the forget set alone is therefore the harder and more realistic regime, where the central difficulty is preserving utility without the remain set to anchor it.



To overcome these limitations, we propose a retain-free unlearning method that updates a pre-trained SER model using only the forget set. Its core idea is to replace the missing remain set with a surrogate synthesised from the model itself: we generate targeted adversarial samples from the forget set as a regularisation signal, and we further constrain each parameter update by its estimated importance so that forgetting does not overwrite general knowledge. Whereas retain-free baselines suppress the forget set but let utility collapse without the remain set, our method (i) drives forget-set performance down to near chance, indicating reduced reliance on the targeted samples, and (ii) retains much of the model's utility on the remaining and unseen-speaker data. On two emotional-speech corpora and two self-supervised backbones, it consistently outperforms retain-free baselines and approaches methods that require the remain set.

\sstitle{Related Work}
Machine unlearning for speech-related tasks has primarily focused on speaker-attribute unlearning and instance-level unlearning. For speaker attribute unlearning, the study in~\cite{reyner2024machine} employed domain-adversarial training to identify gender-based violence victim condition and forget speaker identification information. For instance-level unlearning, the studies in~\cite{cheng2025speech, koudounas2025alexa} employed the remain set during unlearning. The authors in~\cite{cheng2025speech} proposed a forgetting strategy based on curriculum learning to dynamically learn sample weights. However, as discussed above, requiring the remain set can be impractical when it is unavailable or undesirable to use, and it increases storage and computational costs. Negative Gradient (\textit{NegGrad}), which reverses the gradient direction to impair performance on selected samples without using the remain set, has been shown to be an efficient retain-free approach in speech recognition~\cite{koudounas2025alexa}.



In SER, prior work~\cite{bourtoule2021machine, phukan2025towards} has explored shard-based strategies, where multiple models are trained on disjoint data partitions and later combined via weight averaging to enable forgetting. In contrast, we study instance-level unlearning of a single SER model in a retain-free setting that uses only the forget set. This avoids the overhead of training and maintaining multiple SER models and eliminates the need to access the remain set, which can be costly and may introduce additional privacy risks. Inspired by~\cite{cha2024learning}, we build on retain-free baselines including \textit{NegGrad} and Random Labelling (\textit{RandLabel}), where the forget set is assigned random targets during unlearning. We further enhance these baselines by incorporating adversarial-sample-based regularisation and an elastic weight constraint to stabilise updates and better preserve utility.

\section{Methodology}
\label{sec:method}

\begin{figure}[t]
    \centering
    \includegraphics[width=.99\linewidth, trim=0mm 0mm 0mm 6mm, clip]{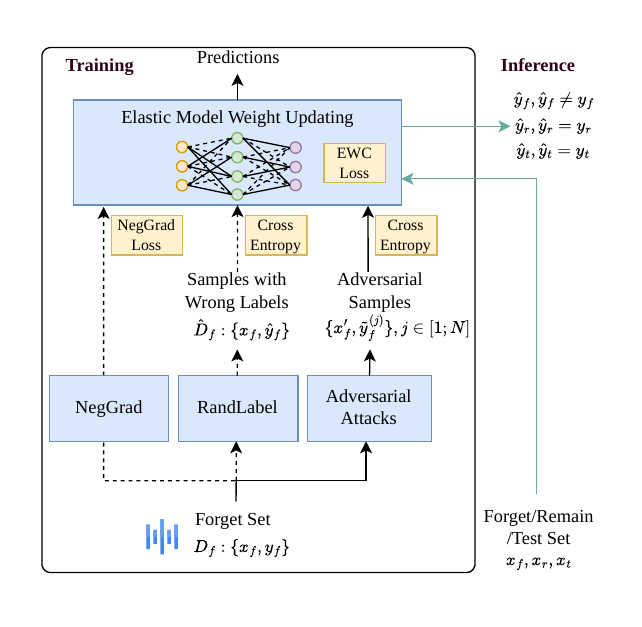}
        \vspace{-25pt}
    \caption{The proposed machine unlearning approach. The dash lines mean to use either \textit{NegGrad} or \textit{RandLabel}.}
    \vspace{-15pt}
    \label{fig:pipeline}
\end{figure}


The training data is denoted by $D:\{\bm{x},y\}$, where $\bm{x}$ is the speech samples and $y$ denotes the emotion labels. We partition $D$ into a forget set $D_f:\{\bm{x_f}, y_f\}$ and a remain set $D_r:\{\bm{x_r}, y_r\}$, such that $D=D_f \cup D_r$ and $D_f \cap D_r = \phi$. Given an SER model $f$ trained on $D$, our goal is to fine-tune $f$ using only $D_f$ so that it (i) degrades performance on $\bm{x_f}$ while (ii) preserving the knowledge associated with $D_r$. As illustrated in Fig.~\ref{fig:pipeline}, forgetting is achieved by \textit{NegGrad} or \textit{RandLabel} (Section~\ref{sec:randomlabel}). Adversarial samples are additionally generated for regularisation. We further employ an elastic model constraint to stabilise updates and better retain model utility on $D_r$ (Section~\ref{sec:rest}).

\subsection{Forgetting via NegGrad and RandLabel}
\label{sec:randomlabel}
We investigate two retain-free objectives to induce forgetting on $\bm{x_f}$, including \textit{NegGrad}~\cite{golatkar2020eternal} and \textit{RandLabel}~\cite{golatkar2020eternal}. \textit{NegGrad} performs gradient ascent on the forget set to reduce the model’s fit to $D_f$, \ie 
\vspace{-5pt}
\begin{equation}
\label{eq:neggrad}
\vspace*{-1mm}    \mathcal{L_{\text{neg}}}=-\mathcal{L_{\text{CE}}}(f(\bm{x_f}), y_f),
\end{equation}
where $\mathcal{L_{\text{CE}}}$ denotes the cross-entropy loss. In \textit{RandLabel}, we assign each forget-set sample a randomly generated incorrect label, yielding $\hat D_f:\{\bm{x_f}, \hat y_f\}$, $\hat y^i_f \neq y^i_f$ for sample index $i$. The model is then optimised to fit these random targets:
\vspace{-5pt}
\begin{equation}
\vspace*{-1mm}    \mathcal{L_{\text{rand}}}=\mathcal{L_{\text{CE}}}(f(\bm{x_f}), \hat{y}_f).
    \label{eq:misclassify}
\end{equation}

\subsection{Regularisation}
\label{sec:rest}
\textit{NegGrad} and \textit{RandLabel} (Section~\ref{sec:randomlabel}) can effectively degrade performance on the forget set. However, unlearning updates may also induce undesired forgetting of knowledge associated with the remain set. To mitigate this, we introduce two regularisation strategies that help preserve model utility on the remain set without requiring access to $D_r$.

\subsubsection{Adversarial Attacks}
Adversarial attacks can craft small, human-imperceptible perturbations that substantially degrade a model’s predictions on the resulting adversarial examples~\cite{ren2020generating}. Adversarial examples have also been explored for data augmentation and can outperform simple perturbations such as random noise~\cite{ren2020generating}. Moreover, prior work suggests that adversarial examples may encode task-relevant features associated with the targeted labels~\cite{ilyas2019adversarial}. Motivated by these findings, we generate adversarial examples and use them as a regulariser, serving as a surrogate signal to help preserve model utility without accessing the remain set.


We generate adversarial samples using targeted Projected Gradient Descent (PGD). For each forget-set sample $x^i_f\in D_f$, we randomly select $M$ target labels $\{\tilde y^{i(1)}_f, \tilde y^{i(2)}_f, ..., \tilde y^{i(M)}_f\}$, where $\tilde y^{i(j)}_f \neq y^{i}_f$. PGD can be viewed as an iterative variant of Fast Gradient Sign Method (FGSM)~\cite{goodfellow2014explaining} performed for $P$ steps~\cite{ren2020generating}. For a given target $\tilde y^{i(j)}_f$, we initialise the attack by adding a small random perturbation bounded by $\tau$: $\tilde x^{i(j)}_{f(1)}=x^{i}_f+\delta^{(j)}$, $|\delta|<\tau$, which encourages diversity among adversarial examples, even when two attacks share the same target label. At the $t$-th step, the adversarial sample is updated as
\vspace{-5pt}
\begin{equation}
\vspace*{-1mm} \tilde x^{i(j)}_{f(t+1)}=\tilde x^{i(j)}_{f(t)}+\sigma*\text{sign}(\nabla\mathcal{L}(x'^{i(j)}_{f(t)}, \tilde y^{i(j)}_f)),
\end{equation}
where $\sigma$ is the step size, $\nabla$ stands for the gradient, and $t=\{1,...,P\}$. After $P$ steps, we limit the difference between adversarial data and real data to be small and invisible with $\tilde x^{i(j)}_{f(P)}=\text{clip}(\tilde x^{i(j)}_{f(P)}, x^{i(j)}_{f(P)}-\tau, x^{i(j)}_{f(P)}+\tau)$. Given the adversarial set and their targets, the model is trained by
\vspace{-5pt}
\begin{equation}
\vspace*{-1mm}    \mathcal{L_{\text{adv}}}=\mathcal{L_{\text{CE}}}(f(\bm{\tilde x_f}), \tilde{y}_f).
\end{equation}

\subsubsection{Elastic Model Weights}
Lifelong learning has been shown to better preserve previously acquired knowledge than standard transfer learning~\cite{ren2020enhancing}. To mitigate unintended forgetting of knowledge associated with $D_r$, we employ Elastic Weight Consolidation (EWC)~\cite{kirkpatrick2017overcoming}, which penalises changes to parameters that are deemed important. Parameter importance is estimated using the (diagonal) Fisher information\cite{desjardins2015natural}, a second-order approximation of the curvature of the loss function. In our retain-free setting, we estimate the Fisher information using gradients of the cross-entropy loss on the forget set, $\mathcal{L}_{\text{CE}}$ ($f(\mathbf{x}_f),y_f$). We then transform it into an inverse square-root weighting to form the diagonal Fisher terms, so that parameters with larger estimated importance receive stronger constraints during unlearning. The resulting EWC penalty is
\vspace{-5pt}
\begin{equation}
\vspace*{-1mm}    \mathcal{L}_{\text{ewc}}=\sum_k F_{k}(\theta_{k}-\theta_{k}^*)^2,
\end{equation}
where $\theta$ denotes the parameters of $f$, $\theta^*$ are the initial parameters, and $F_k$ is the $k$-th diagonal entry of the Fisher matrix.


\subsubsection{Model Training}
\label{sec:training}
We optimise the model to induce forgetting on $D_f$ and preserve utility associated with $D_r$ via a weighted combination of losses:
\vspace{-5pt}
\begin{equation} 
\mathcal{L}=\lambda_1\mathcal{L}_{\text{mis}}+\lambda_2\mathcal{L}_{\text{adv}}+\lambda_3\mathcal{L}_{\text{ewc}},
\vspace*{-1mm}
\end{equation}
where $\mathcal{L}_{\text{mis}}\in\{\mathcal{L}_{\text{neg}},\,\mathcal{L}_{\text{rand}}\}$, and $\lambda_{1},\lambda_{2},\lambda_{3}$ are scalar coefficients controlling the contributions of each term.

\begin{table*}[t]
    \centering
    \caption{Model performance (UAR) on $D_f$, $D_r$ and the test set $D_t$, when forgetting $N$ samples. The best performance of the proposed approaches are compared to baselines (*: $p<0.001$ in a one-tailed z-test). }
    \scalebox{0.78}{
    \begin{tabular}{l|p{0.95cm}p{0.95cm}p{0.95cm}|p{0.95cm}p{0.95cm}p{0.95cm}|p{0.95cm}p{0.95cm}p{0.95cm}|p{0.95cm}p{0.95cm}p{0.95cm}}
    \toprule
        & \multicolumn{3}{c|}{$N=10$} & \multicolumn{3}{c|}{$N=30$} & \multicolumn{3}{c|}{$N=50$} & \multicolumn{3}{c}{$N=100$}\\
        \hline
         & $D_f\downarrow$ & $D_r\uparrow$ & \textit{$D_t\uparrow$}  & $D_f\downarrow$ & $D_r\uparrow$& \textit{$D_t\uparrow$} & $D_f\downarrow$ & $D_r\uparrow$& \textit{$D_t\uparrow$} & $D_f\downarrow$ & $D_r\uparrow$& \textit{$D_t\uparrow$}  \\
        \hline
        Original (all) & 0.857 & 0.996 & 0.902 &1.000 &0.995 &0.902 &1.000 &0.995 &0.902  &1.000 &0.995 &0.902  \\
        Original ($D_r$) &0.857 &0.997 & 0.907  &0.964 &0.998 &0.905  & 0.982 & 0.995 & 0.901 &0.963 &0.988&0.891 \\        
        \hline
        \multicolumn{13}{c}{\textit{Upper bound: Machine unlearning with forget set $D_f$ and remain set $D_r$}}\\        
        \hline
        UNSIR~\cite{tarun2023fast} &1.000 &0.978 &0.893 &0.964 &0.968 &0.890  &0.966 &0.954 &0.855  &0.920 &0.941 &0.861 \\
        Bad Teaching~\cite{chundawat2023can} &0.143 &0.974 &0.860 &0.429 &\textbf{0.995} &\textbf{0.906} &0.650 &0.994 &\textbf{0.905} &0.154 &0.993 &0.902  \\
        SCRUB~\cite{zhang2024towards} &0.643 &\textbf{0.979} &\textbf{0.888}  &0.286 &0.962 & 0.863 &0.775 &0.975 &0.891 & 0.143 & 0.981 & 0.878  \\        
        NegGrad+~\cite{choi2023towards, kurmanji2023towards} & \textbf{0.000} &0.975 &0.872 &0.129 &0.981 & 0.894 &0.130 &\textbf{0.996} & 0.902 &0.132 &\textbf{0.995} & \textbf{0.906} \\
        RandLabel+~\cite{cheng2025speech} &\textbf{0.000} &0.923 &0.824  &\textbf{0.119} & 0.964 &0.852  &\textbf{0.065} & 0.980 & 0.882  & \textbf{0.106} & 0.984 & 0.884  \\        
        \hline
        \multicolumn{13}{c}{\textit{Proposed setting: Machine unlearning with forget set $D_f$ only}}\\
        \hline        
        NegGrad~\cite{golatkar2020eternal} (Baseline) & 0.071 & 0.834 & 0.726  & 0.143 &0.218 & 0.228 &0.118 &0.202&0.212 &\textbf{0.106} &0.210&0.197 \\
        \textbf{NegGrad-Adv} & 0.071 & 0.951 & 0.859  & 0.136 & 0.851 & 0.757  & \textbf{0.097} & 0.651 & 0.558 &0.302 & 0.399 & 0.360 \\
        \textbf{NegGrad-Adv-EWC} & \textbf{0.000} & \textbf{0.985*} & \textbf{0.886*} & \textbf{0.129} & \textbf{0.948*} & \textbf{0.849*} & 0.124 & \textbf{0.860*} &\textbf{0.761*} & 0.111 & \textbf{0.586*} & \textbf{0.513*} \\              
        \hline 
        RandLabel~\cite{golatkar2020eternal} (Baseline) & \textbf{0.000} & 0.735 & 0.672 & 0.083 & 0.744 & 0.664  & 0.136 &0.646 & 0.572 & \textbf{0.104} &0.538 & 0.475 \\            
        \textbf{RandLabel-Adv} & \textbf{0.000} &0.940 & 0.836 & 0.093 & 0.864 & 0.766  &0.081 &0.755 &0.665  &0.122 &0.775 & 0.678 \\
        \textbf{RandLabel-Adv-EWC} &\textbf{0.000} & \textbf{0.968*} & \textbf{0.871*}  & \textbf{0.036} & \textbf{0.943*} & \textbf{0.844*}  & \textbf{0.063} & \textbf{0.846*} & \textbf{0.750*}  &0.139 & \textbf{0.791*} & \textbf{0.702*}  \\        
    \bottomrule
    \end{tabular}
    }
    \label{tab:result}
\end{table*}

\section{Experimental Results}
\label{sec:experiment}

\subsection{Databases}
\label{sec:data}
The DEMoS corpus~\cite{parada2020demos} is an Italian emotional speech dataset recorded from $68$ speakers ($23$ females and $45$ males).
It contains $9\,365$ utterances annotated with seven classes:
\textit{Anger}, \textit{Disgust}, \textit{Fear}, \textit{Guilt}, \textit{Happiness}, \textit{Sadness}, and \textit{Surprise}. We use a speaker-independent split with 
$3\,024$/$3\,024$/$3\,317$ utterances for training/validation/test, respectively. Detailed class distributions are provided in~\cite{ren2020generating, ren2020enhancing}.


The IEMOCAP corpus~\cite{busso2008iemocap} is an English emotional speech dataset comprising $10$ speakers across five sessions. Following common practice, we merge \textit{happiness} and \textit{excitement} and consider four classes: \textit{anger}, \textit{happiness}, \textit{neutral}, and \textit{sadness}. We adopt a speaker-independent split into $3\,259$/$1\,031$/$1\,241$ utterances for training/validation/test.


\begin{figure}[!htp]
    \centering
    \includegraphics[width=.9\linewidth]{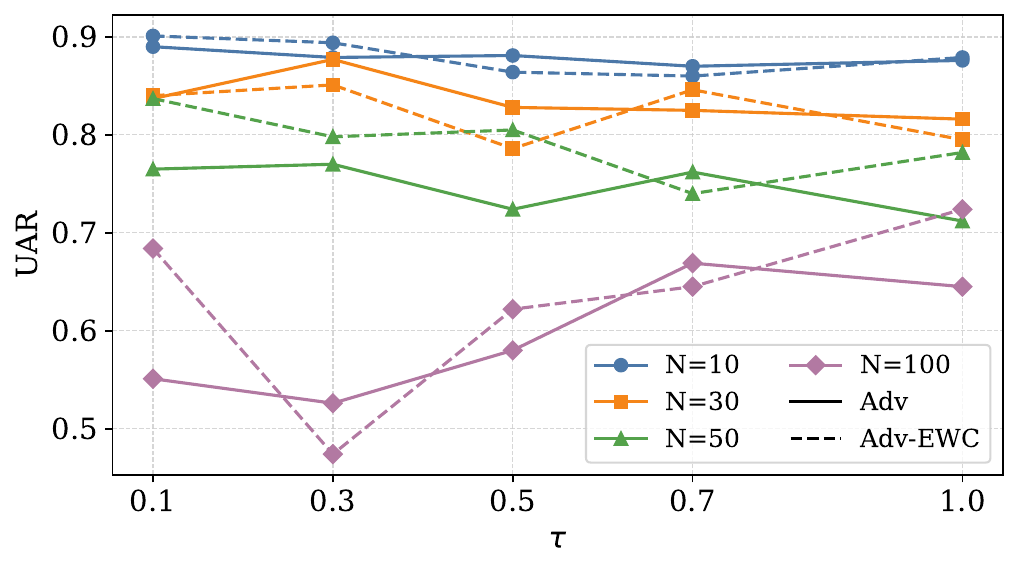}
    \vspace{-12pt}
    \caption{Validation performance of unlearning models with RandLabel+adversarial samples when forgetting $N$ samples.
    }
    \vspace{-10pt}
    \label{fig:val}
\end{figure}

\subsection{Experimental Setup}
\label{sec:expset}


We fine-tune pretrained Wav2Vec 2.0 or HuBERT models~\cite{baevski2020wav2vec, hsu2021hubert} (trained on LibriSpeech~\cite{panayotov2015librispeech}) on $16$\,kHz resampled speech for 20 epochs using Adam, with learning rate $3\times10^{-5}$ and batch size 16. The resulting model (\textit{Original (all)}) is then unlearned for up to 20 epochs with early stopping once the UAR on $D_f$ falls below chance. We generate $M=20$ adversarial examples per forget sample using targeted PGD with $P=50$ steps. Loss weights are set to $\lambda_1=0.1$, $\lambda_2=1$, and $\lambda_3=0.01$ for stable optimisation. For \textit{NegGrad} with regularisation, we additionally penalise the probability of the ground-truth class with weight $0.05$.
We report Unweighted Average Recall (UAR) to quantify forgetting on $D_f$ and utility on $D_r$ as well as on unseen-speaker validation and test sets.


\sstitle{Baselines and State-of-the-Art} 
We use \textit{NegGrad} and \textit{RandLabel} (Section~\ref{sec:randomlabel}) as retain-free baselines. For \textit{NegGrad}, we minimise negative entropy and the probability of the ground-truth class to improve training stability.


Besides \textit{Original (all)}, we also fine-tune a model on the remain set only, denoted \textit{Original ($D_r$)}. We further include the following state-of-the-art methods that use both $D_f$ and $D_r$ as upper bounds.
(i) \textit{UNSIR}\cite{tarun2023fast} generates error-maximising noise for $D_f$ and retrains on the noise combined with $D_r$.
(ii) \textit{Bad Teaching}\cite{chundawat2023can} trains a student to match \textit{Original (all)} on $D_r$ and a random predictor on $D_f$.
(iii) \textit{SCRUB}\cite{zhang2024towards} trains a student to mimic \textit{Original (all)} on $D_r$ while deviating on $D_f$.
(iv) \textit{NegGrad+}\cite{choi2023towards, kurmanji2023towards} improves \textit{NegGrad} by additionally training on $D_r$ to recover utility.
(v) \textit{RandLabel+}~\cite{golatkar2020eternal} augments \textit{RandLabel} with additional training on $D_r$.


\subsection{Performance Comparison}
\label{sec:result}

Figure~\ref{fig:val} reports validation performance when forgetting $N\in[10,100]$ samples and varying the PGD clipping bound $\tau\in[0.1,1.0]$. We compare two variants: adversarial regularisation (\textit{Adv}) and adversarial regularisation with elastic weight constraints (\textit{Adv-EWC}). Overall, performance decreases slightly as $\tau$ increases, likely because a larger perturbation budget introduces a stronger distribution shift from the original data. Performance is generally higher for smaller $N$, as forgetting becomes more difficult when more samples are removed. For $N=100$, results are less stable than for $N\in\{10,30,50\}$, possibly because the generated adversarial targets occasionally exhibit class-distribution mismatch relative to the remaining data when $N$ is large. Based on Fig.~\ref{fig:val}, we set $\tau=0.7$ for \textit{Adv} and $\tau=0.1$ for \textit{Adv-EWC} in subsequent experiments.

Table~\ref{tab:result} shows that \textit{Original ($D_r$)} achieves performance comparable to \textit{Original (all)} on $D_f$, $D_r$, and the test set $D_t$. Unlearning methods that leverage $D_r$ generally maintain utility close to the original models, as $D_r$ provides abundant real training data. Retain-free methods trained only on $D_f$ typically yield lower utility, especially when $N=50,100$, since unlearning is performed with substantially less data. Notably, some $D_r$-based methods (\eg UNSIR, Bad Teaching, and SCRUB) obtain above-chance performance on $D_f$ (chance=$0.143$ for seven-class classification), indicating incomplete forgetting of the targeted samples.



Compared with the retain-free baselines, our \textit{Adv} and \textit{Adv-EWC} variants more effectively reduce the model’s reliance on $D_f$ (UAR $\leq 0.139$) with the exception of \textit{NegGrad-Adv}. \textit{RandLabel} variants are more stable than \textit{NegGrad}, as they optimise standard cross-entropy rather than gradient ascent. At the same time, the proposed methods consistently improve utility relative to the baselines, highlighting the benefit of adversarial samples as a regularisation/augmentation signal in the absence of $D_r$.
Adding EWC further improves over \textit{Adv}, likely by constraining updates to parameters that are less critical for the original model, thereby stabilising unlearning. As $N$ increases, performance degrades for all retain-free methods, potentially due to stronger distribution shift induced by a larger set of generated adversarial examples. Nevertheless, our methods still yield significant gains over the baselines ($p<0.001$ in a one-tailed z-test).

\subsection{Ablation Study}
\label{sec:ablation}         

Table~\ref{tab:result_model_demos} reports results on DEMoS when varying the number of forgotten speakers, using both Wav2Vec 2.0 and HuBERT~\cite{hsu2021hubert}. Across backbones and speaker settings, the proposed \textit{Adv} and \textit{Adv-EWC} variants (built on either \textit{NegGrad} or \textit{RandLabel}) generally achieve significantly better performance than the retain-free baselines, further supporting the effectiveness of our regularisation strategy. The best overall performance is achieved by \textit{RandLabel-Adv-EWC}. This is consistent with the observation that \textit{RandLabel} is typically more stable than \textit{NegGrad}, as it optimises a standard cross-entropy objective with randomly assigned targets, whereas \textit{NegGrad} relies on gradient ascent. We also observe similar performance when forgetting the same number of samples from one speaker versus two speakers. We further validate the proposed methods on IEMOCAP (Table~\ref{tab:result_model_iemocap}). Consistent with DEMoS, \textit{Adv-EWC} yields the strongest performance among retain-free methods and outperforms both the baselines and \textit{Adv} for unlearning under both \textit{NegGrad} and \textit{RandLabel}.

\subsection{Visualisation of Speech Representations}

\begin{figure}[!t]
    \centering
    \includegraphics[width=\linewidth, trim=7.5mm 7mm 6.5mm 6mm, clip]{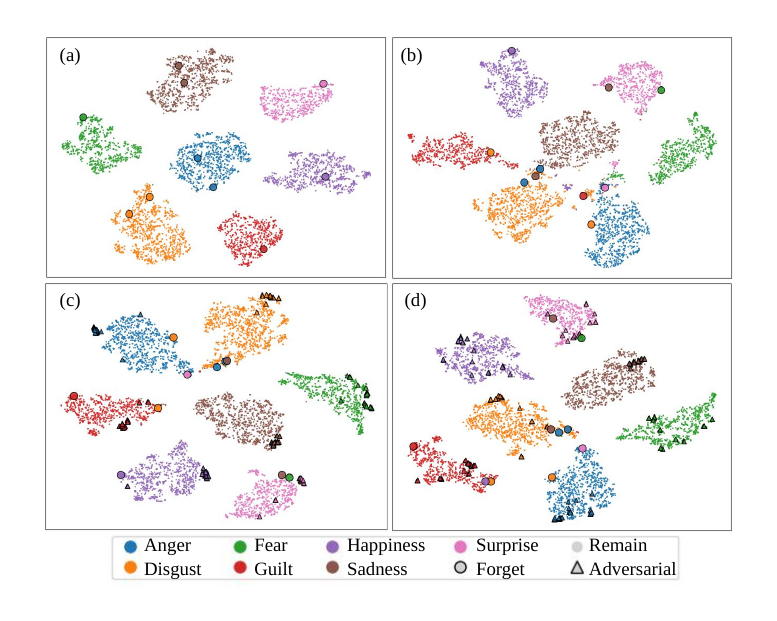}
    \vspace{-18pt}
    \caption{T-SNE visualisation of representations from the linear layer following wav2vec 2.0 when forgetting 10 samples from DEMoS. (a) Original (all), (b) RandLabel, (c) RandLabel-Adv, (d) RandLabel-Adv-EWC.}
    \label{fig:tsne}
\end{figure}

We visualise the learnt representations of $D_f$, $D_r$, and the generated adversarial samples, extracted from the penultimate layer of the four models, in Fig.~\ref{fig:tsne} to better understand why adversarial regularisation is effective. Compared with \textit{Original (all)}, \textit{RandLabel} drives $D_f$ toward misclassification while slightly degrading performance on $D_r$. Notably, the adversarial samples largely overlap with the per-class clusters, suggesting that they provide a surrogate signal that captures aspects of the remaining data distribution. Although \textit{RandLabel-Adv} and \textit{RandLabel-Adv-EWC} use the same adversarial samples, \textit{RandLabel-Adv-EWC} yields more diverse within-class representations, which better cover the class structure and may explain its improved utility.

\begin{table}[]
    \centering
    \caption{Ablation study on DEMoS (UAR). }
    \label{tab:result_model_demos}
    \vspace{-10pt}
    \scalebox{0.78}{
    \begin{tabular}{l|p{0.7cm}p{0.7cm}p{0.7cm}|p{0.7cm}p{0.7cm}p{0.7cm}}
    \toprule
    & \multicolumn{3}{c|}{Wav2Vec 2.0}  & \multicolumn{3}{c}{HuBERT}\\
    \hline
    \textit{Method} & $D_f\downarrow$ &$D_r\uparrow$& $D_t\uparrow$ & $D_f\downarrow$ & $D_r\uparrow$ & $D_t\uparrow$  \\
    \hline
      Original (all)  & -- & --& 0.902 & -- & --& 0.902   \\
      \hline
      \multicolumn{7}{c}{\emph{$N=30$, 1 speaker removed}}\\
      \hline
      NegGrad+~\cite{kurmanji2023towards}   & 0.129 & 0.981 & 0.894 &0.171 &0.974 &0.863  \\
      RandLabel+~\cite{cheng2025speech}   & 0.119 & 0.964 & 0.852 &0.093 &0.961 &0.845\\
      \hline
      NegGrad~\cite{golatkar2020eternal}   & 0.143 &0.218 & 0.228 &0.171 &0.207 &0.191  \\
      RandLabel~\cite{golatkar2020eternal}   & 0.083 & 0.744 & 0.664  & 0.121 &0.668& 0.589 \\
      \textbf{NegGrad-Adv}   & 0.136 & 0.851 & 0.757 & 0.083 &0.775 & 0.658   \\
      \textbf{RandLabel-Adv}   & 0.093 & 0.864 & 0.766 & 0.131 &0.896& 0.797 \\      
      \textbf{NegGrad-Adv-EWC}   & 0.129 & \textbf{0.948*} & \textbf{0.849*} &\textbf{0.071}&0.931 &0.808 \\ 
      \textbf{RandLabel-Adv-EWC}   & \textbf{0.036} &0.943 & 0.844 &0.131 &\textbf{0.966*}&\textbf{0.850*} \\ 
          \hline
      \multicolumn{7}{c}{\emph{$N=30$, 2 speakers removed}}\\
      \hline
      NegGrad+~\cite{kurmanji2023towards}  & 0.155 & 0.993 &0.903 &0.107 &0.978 &0.868 \\
      RandLabel+~\cite{cheng2025speech}   & 0.060 & 0.915 &0.807 &0.083 &0.930 &0.830 \\
      \hline
      NegGrad~\cite{golatkar2020eternal}   & 0.119 & 0.449 &0.412 &0.286 &0.444 &0.402 \\
      RandLabel~\cite{golatkar2020eternal}   & 0.136 & 0.658  & 0.596 & 0.136 & 0.547 & 0.482 \\
      \textbf{NegGrad-Adv}   & 0.095 & 0.899 & 0.800 & 0.083 &0.685 & 0.595   \\
      \textbf{RandLabel-Adv}   & 0.071 & 0.875 & 0.772 & 0.083 &0.919 & 0.815  \\      
      \textbf{NegGrad-Adv-EWC}   & 0.119 & 0.939 &0.837 & \textbf{0.100} & 0.924 & 0.803 \\ 
      \textbf{RandLabel-Adv-EWC}   & \textbf{0.048} & \textbf{0.956*} &\textbf{0.857*} & 0.107 & \textbf{0.948*} & \textbf{0.829*} \\ 
    \hline
      \multicolumn{7}{c}{\emph{$N=100$, 1 speaker removed}}\\     \hline
      NegGrad+~\cite{kurmanji2023towards}   & 0.132 &0.995& 0.906 &0.197 &0.994 &0.894  \\
      RandLabel+~\cite{cheng2025speech}   & 0.106 & 0.984& 0.884 &0.072 &0.972&0.855 \\
      \hline
      NegGrad~\cite{golatkar2020eternal}   & \textbf{0.106} &0.210& 0.197 &0.135 &0.205&0.197\\
      RandLabel~\cite{golatkar2020eternal}   & 0.104 &0.538 & 0.475  & 0.139 &0.477& 0.408 \\
      \textbf{NegGrad-Adv}   & 0.302 &0.399& 0.360 & 0.135 & 0.366& 0.317  \\
      \textbf{RandLabel-Adv}   & 0.122 &0.775& 0.678 & 0.120 &0.541& 0.457 \\      
      \textbf{NegGrad-Adv-EWC}   & 0.111 &0.586& 0.513 &0.139 &0.416 &0.338 \\ 
      \textbf{RandLabel-Adv-EWC}   & 0.139 &\textbf{0.791*}& \textbf{0.702*} &\textbf{0.118} & \textbf{0.807*} &\textbf{0.672*} \\ 
          \hline
      \multicolumn{7}{c}{\emph{$N=100$, 2 speakers removed}}\\
      \hline
      NegGrad+~\cite{kurmanji2023towards}   & 0.119 & 0.998 &0.909 &0.191 &0.977 &0.854 \\
      RandLabel+~\cite{cheng2025speech}   & 0.059 & 0.980 &0.880 &0.103 &0.945 &0.826 \\
      \hline
      NegGrad~\cite{golatkar2020eternal}   & 0.143 & 0.163 &0.161 &0.278 &0.342 &0.289 \\
      RandLabel~\cite{golatkar2020eternal}   & 0.127 & 0.370  & 0.336 & 0.139 & 0.473 & 0.412 \\
      \textbf{NegGrad-Adv}   & 0.112 & 0.415 & 0.366 & 0.155 &0.227 & 0.204  \\
      \textbf{RandLabel-Adv}   & 0.127 & 0.620 & 0.527 & \textbf{0.064} & 0.533 & 0.451  \\      
      \textbf{NegGrad-Adv-EWC}   & 0.125 & 0.455 &0.410 & 0.226 & 0.288 & 0.245 \\ 
      \textbf{RandLabel-Adv-EWC}   & \textbf{0.040} & \textbf{0.806*} &\textbf{0.708*} & 0.124 & \textbf{0.766*} & \textbf{0.646*} \\       
    \bottomrule
    \end{tabular}}
\end{table}
\vspace{-10pt}

\begin{table}[]
    \centering
    \caption{Model comparison on IEMOCAP (UAR). }
    \label{tab:result_model_iemocap}
    \vspace{-10pt}
    \scalebox{0.78}{
    \begin{tabular}{l|p{0.7cm}p{0.7cm}p{0.7cm}|p{0.7cm}p{0.7cm}p{0.7cm}}
    \toprule
    & \multicolumn{3}{c|}{Wav2Vec 2.0}  & \multicolumn{3}{c}{HuBERT} \\
    \hline
    \textit{Method} & $D_f\downarrow$ & $D_r\uparrow$ & $D_t\uparrow$ & $D_f\downarrow$ & $D_r\uparrow$ & $D_t\uparrow$  \\
    \hline
      Original (all)  & -- & -- & 0.659 & -- & -- & 0.677  \\
      \hline
      \multicolumn{7}{c}{\emph{$N=30$, 2 speakers removed}}\\
      \hline
      NegGrad+~\cite{kurmanji2023towards}   & 0.225 & 0.990 &0.687 &0.168 &0.764 &0.516 \\
      RandLabel+~\cite{cheng2025speech}   & 0.239 & 0.948 &0.633 &0.210 &0.847 &0.557 \\
      \hline
      NegGrad~\cite{golatkar2020eternal}   & 0.420 & 0.472 &0.416 &0.205 &0.406 &0.312 \\
      RandLabel~\cite{golatkar2020eternal}   & 0.167 & 0.520  & 0.391 & 0.237 & 0.797 & 0.520 \\
      \textbf{NegGrad-Adv}   & \textbf{0.064} & 0.730 & 0.499 & \textbf{0.163} &0.788 & 0.510  \\
      \textbf{RandLabel-Adv}   & 0.167 & 0.792 & 0.510 & 0.227 &0.825 & 0.524  \\      
      \textbf{NegGrad-Adv-EWC}   & 0.224 & \textbf{0.812*} &0.549 & 0.244 & 0.772 & 0.496 \\ 
      \textbf{RandLabel-Adv-EWC}   & 0.247 & 0.806 &\textbf{0.578*} &0.185 & \textbf{0.864*} & \textbf{0.546} \\ 
    \hline
      \multicolumn{7}{c}{\emph{$N=100$, 2 speakers removed}}\\
      \hline
      NegGrad+~\cite{kurmanji2023towards}   & 0.219 & 0.988 &0.659 &0.143 &0.972 &0.630 \\
      RandLabel+~\cite{cheng2025speech}   & 0.224 & 0.948 &0.614 &0.214 &0.921 &0.575 \\
      \hline
      NegGrad~\cite{golatkar2020eternal}   & 0.363 & 0.390 &0.335 &0.238 &0.251 &0.259 \\
      RandLabel~\cite{golatkar2020eternal}   & 0.220 & 0.228  & 0.202 & 0.245 & 0.318 & 0.270 \\
      \textbf{NegGrad-Adv}   & \textbf{0.118} & 0.461 & 0.301 & 0.198 &0.671 & 0.430  \\
      \textbf{RandLabel-Adv}   & 0.240 & 0.683 & 0.460 & \textbf{0.121} &0.676 & 0.419  \\      
      \textbf{NegGrad-Adv-EWC}   & 0.189 & 0.581 &0.385 & 0.246 & \textbf{0.774*} & \textbf{0.495*} \\ 
      \textbf{RandLabel-Adv-EWC}   & 0.247 & \textbf{0.729*} &\textbf{0.497*} & 0.157 & 0.765 & 0.458 \\ 
    \bottomrule
    \end{tabular}}
\end{table}

\section{Conclusion and Future Work}
\label{sec:conclude}

This work presented a retain-free machine unlearning approach for SER that used only the forget set to mitigate privacy risks in emotional speech. Our method generated targeted adversarial examples as a regularisation/augmentation signal and further stabilised unlearning by constraining parameter updates according to their estimated importance. Experiments showed that the proposed approach drove performance on the forget set toward chance while largely preserving utility on the remain set and on unseen-speaker validation/test data.
We also find that increasing the size of the forget set degrades utility, suggesting that more principled adversarial-generation strategies (\eg prototype- or cluster-based adversarial examples that better match the remaining distribution) are needed for large-scale deletions. Future work will extend retain-free unlearning to structured requests, such as forgetting specific emotion classes and sensitive attributes (\eg gender).

\bibliographystyle{IEEEtran}
\bibliography{mybib}

\end{document}